\documentclass[twocolumn]{aastex701}
\usepackage{subcaption}
\usepackage{amsmath}

\newcommand\padiffdustjet{| \Psi_{\rm dust - jet} |}
\newcommand\padiffgaljet{| \Psi_{\rm gal - jet} |}
\newcommand\padiffgaldust{| \Psi_{\rm gal - dust} |}

\begin{document}

\title{The Perpendicularity of Dust Lanes and Radio Jets in Early-Type Galaxies:\\Implications for AGN Feedback}

\author[orcid=0009-0000-6612-0599, gname='Emma', sname='Weller']{Emma Jane Weller}
\affiliation{Department of Astronomy, Yale University, New Haven, CT 06511, USA}
\email[show]{emma.weller@yale.edu}

\author[orcid=0000-0002-8282-9888, gname='Pieter', sname='van Dokkum']{Pieter van Dokkum}
\affiliation{Department of Astronomy, Yale University, New Haven, CT 06511, USA}
\email{pieter.vandokkum@yale.edu}

\begin{abstract}
The orientation of radio jets relative to their host galaxies offers an interesting avenue for probing the connection between active galactic nuclei (AGN) and their surroundings. Several studies have also investigated the orientation of nuclear dust features. We follow up on this previous work with newer Hubble Space Telescope imaging of early-type radio galaxies, and a largely automated process for measuring position angles. We classify the dust features as lanes, disks, or rings. Lanes are irregular structures that likely form from gas-rich minor mergers, while disks and rings are more well-defined and may form from settling lanes or internal mechanisms. We find that dust lanes do not have a preferred alignment relative to their host galaxies, but are preferentially perpendicular to the jets. In contrast, dust disks and rings tend to be closely aligned with the major axes of their host galaxies, but have varying orientations relative to the jets. Our results suggest that infalling dusty material from mergers can influence the angle of the radio jet. This would allow the jet orientation to change over time, and may help explain the role of AGN feedback in maintaining quiescence in massive galaxies.
\end{abstract}


\section{Introduction} \label{sec:intro}

It is well known that active galactic nuclei (AGN) have significant impacts on the evolution of their host galaxies. They are believed to help quench star formation and maintain quiescence by heating and ejecting gas from the galactic center (e.g. \citealt{Silk_Rees_1998, DiMatteo+2005, Springel+2005, Hopkins+2006, Schaye+2015, Weinberger+2018, Weller+2025}). One way to investigate the interaction between central supermassive black hole (SMBH) accretion and the surrounding environment is through AGN radio jets. The connection between SMBH spin, the inner accretion disk, and larger-scale structures and inflows has been studied theoretically (e.g. \citealt{Natarajan_Pringle_1998, Hopkins+2012, McKinney+2013}). Observationally, many studies have looked for a preferred alignment between radio jets (over a range of scales) and their host galaxies, with varying results: some found that the jets are preferentially perpendicular to the major axis of the host galaxy, while others concluded that the jet orientation is random (e.g. \citealt{Gibson_1975, Sullivan_Sinn_1975, Guthrie_1979, Palimaka+1979, Birkinshaw_Davies_1985, Sansom+1987, Condon_1991, Battye_Browne_2009, Zheng+2024, FernandezGil+2025}). In a study using data from multiple radio and optical surveys, \cite{Zheng+2024} proposed that in most radio galaxies, the central SMBH accretes in a coherent manner, usually resulting in a jet approximately perpendicular to the host galaxy's major axis. However, they suggested that many of the luminous radio AGN with massive hosts experience chaotic accretion or galaxy mergers, and therefore have randomly oriented jets.

The idea that mergers drive gas into the galactic center and trigger a burst of star formation and AGN activity has been well-studied in simulations (e.g. \citealt{Hernquist_1989, DiMatteo+2005, Springel+2005, Hopkins+2006, Hopkins+2008, Pontzen+2017}). Dust lanes in early-type galaxies can serve as a tracer of recent gas-rich minor mergers. Using the Sloan Digital Sky Survey, \cite{Kaviraj+2012} and \cite{Shabala+2012} found several indicators of a merger origin for these lanes, and concluded that their hosts represent a phase of starburst galaxy evolution in which AGN activity has been triggered but has not yet completely quenched star formation. \cite{vanDokkum_Franx_1995} and \cite{VerdoesKleijn_deZeeuw_2005} (hereafter VK05) found that the dust detection rate is higher in early-type galaxies with radio jets than in those without jets.

In an early study of seven elliptical radio galaxies crossed by a band of dust absorption, \cite{Kotanyi_Ekers_1979} found that in all cases the jets were ``nearly perpendicular'' to the dust bands, with position angle (PA) differences of $60^\circ - 90^\circ$. With the launch of the Hubble Space Telescope (HST), it became possible to routinely detect nuclear dust disks in nearby galaxies. These disks are more compact than dust lanes, and they may be late, relatively stable stages of minor accretion events (see \citealt{vanDokkum_Franx_1995}). \cite{Jaffe+1993} used HST to identify a disk of cool dust and gas perpendicular to the radio jets in the elliptical galaxy NGC 4261. Several studies further investigated the alignment of jets and dust features (both lanes and disks) using HST imaging, with sample sizes ranging from a few galaxies to over $30$ (e.g., \citealt{vanDokkum_Franx_1995, VerdoesKleijn+1999, deKoff+2000, Martel+2000, deRuiter+2002, Schmitt+2002}; VK05). As with the studies of galaxy-jet alignment, the results varied. Some agreed with \cite{Kotanyi_Ekers_1979}, while \cite{Schmitt+2002} found no preferentially perpendicular alignment. VK05 differentiated between dust ``ellipses,'' which they interpreted as thin, nearly circular, settled disks, and dust lanes, which they suggested are warped structures that may be in the process of settling into disks. They found that (i) dust ellipses are preferentially aligned with the major axes of their host galaxies, while dust lanes typically show no preferred alignment; (ii) jets have no preferred alignment with respect to their host galaxies (and by extension, to dust ellipses); and (iii) jets are preferentially perpendicular to dust lanes.

In this work, we revisit the question of galaxy-dust-jet alignment in early-type radio galaxies, using newer HST imaging where available, and a mostly automated process for measuring PAs. In Sec. \ref{sec:data}, we describe the sources of our images and our galaxy sample. In Sec. \ref{sec:methods}, we explain our process for measuring the galaxy, dust, and jet PAs. In Sec. \ref{sec:results}, we present an image gallery of the galaxies in our sample, as well as a table of PA measurements from this work and previous studies. In Sec. \ref{sec:analysis}, we analyze our results, and in Sec. \ref{sec:discussion}, we discuss their significance. Finally, in Sec. \ref{sec:conclusions}, we summarize our findings and conclude the paper.

\section{Data} \label{sec:data}

\subsection{Images} \label{sec:images}

We used optical images from HST, accessed using the Mikulski Archive for Space Telescopes (MAST) via the \texttt{Astroquery} package \citep{Ginsburg_2019}. Specifically, we used the High Resolution and Wide Field Channels of the Advanced Camera for Surveys (ACS/HRC and ACS/WFC), or, if suitable images were not available from ACS, the Planetary Camera of the Wide Field and Planetary Camera 2 (WFPC2/PC). We used wide-band filters with effective wavelengths ranging from $435-814 \, {\rm nm}$. For a given galaxy name, HST instrument, and filter, \texttt{Astroquery} returned a table of available science images. With one exception (NGC 5141, noted in Fig. \ref{fig:Obs_gallery_4}), we used the first image in the observations table. The data may be obtained from MAST at \dataset[doi:10.17909/q62n-w257]{https://dx.doi.org/10.17909/q62n-w257}.

We used radio data from the National Radio Astronomy Observatory (NRAO) Very Large Array (VLA) Sky Survey (NVSS, \citealt{Condon+1998}) and the VLA Faint Images of the Radio Sky at Twenty-Centimeters survey (FIRST, \citealt{Becker+1995}). We accessed these data using NASA's \textit{SkyView} virtual observatory \citep{McGlynn+1998}. For each galaxy, we centered our search on the coordinates of the galaxy center as identified in our fitting procedure (see Sec. \ref{sec:gal_dust_PAs}). We chose to use these surveys, rather than taking radio jet PAs from the literature as many previous works have done, in order to have a consistent measurement method. We used NVSS images for more extended emission and FIRST images (when available) for smaller structures that appeared as blobs in NVSS due to the lower resolution.

\subsection{Galaxy sample} \label{sec:sample}

Our sample of early-type radio galaxies with nuclear dust was assembled primarily from galaxies used in prior related studies. We identified additional potential sample members using the NASA/IPAC Extragalactic Database (NED) Search for Objects by Classifications tool. We chose galaxies for our sample if they had available imaging by ACS/HRC, ACS/WFC, or WFPC2/PC and by NVSS or FIRST, and if there was clearly visible, approximately nuclear dust absorption. Galaxies with face-on dust disks or very diffuse or patchy dust were excluded due to the difficulty of measuring a dust PA. Galaxies were also excluded if their jets were too compact and unresolved to measure a jet PA. This resulted in a sample of 32 galaxies, ranging from $z \sim 0.003 - 0.1$. The nuclear dust features of all these galaxies have been reported in the literature (\citealt{Kotanyi_Ekers_1979, vanDokkum_Franx_1995, Capetti+2000, deKoff+2000, Martel+2000, Schmitt+2002}; VK05; \citealt{Schmitt+2002, Davidson_2024}).

\section{Methods} \label{sec:methods}

\subsection{Measuring galaxy and dust PAs} \label{sec:gal_dust_PAs}

Our optical image processing procedure was developed by trial and error on a subset of the galaxies in our sample. We began by calculating the 99.99th percentile pixel brightness in the image and taking this to be the maximum value. We set all pixels above the threshold to this maximum, and all pixels with NaN or negative brightness values to zero. Next, we identified sources in the image using the \texttt{Photutils} package \citep{Bradley+2025}, with a threshold of 5 standard deviations above the median brightness and a minimum size of 100 pixels. We took the brightest source to be the galaxy, though occasionally we had to exclude a portion of the image containing a bright spot or defect in order for the galaxy to be correctly identified (these cases are noted in Fig. \ref{fig:Obs_gallery_1}). We used \texttt{Photutils} to calculate the center of the source and the PA of the major axis, which we mark with a blue line in Fig. \ref{fig:Obs_gallery_1}.

Next, we attempted to identify the dust automatically by creating a Gaussian model for the elliptical galaxy fit with $\sigma = 20 \, {\rm pix}$, subtracting it out, and inverting the image. We calculated the 20th percentile pixel brightness and took this to be the minimum value, setting all pixels below the threshold to zero. Using the same process as above, except with a threshold of 10 standard deviations, we identified the brightest source in the inverted image using \texttt{Photutils}, took this to be the dust feature, and extracted the center and PA. Unlike the galaxy fitting, however, this process often failed to match the PA of the visually apparent dust feature, due to the nonuniform nature of the dust. In these cases, we fit the center and PA manually. We show the measured dust PAs with red lines in Fig. \ref{fig:Obs_gallery_1}.

\subsection{Measuring radio jet PAs} \label{sec:radio_PAs}

Our radio image processing procedure was also developed by trial and error. We began by identifying all the radio sources in the image using the Python Blob Detector and Source Finder package (\texttt{PyBDSF}, \citealt{Mohan_Rafferty_2015}). Then, we considered the 9 brightest sources (or all the sources if there were less than 9), and took the one that was closest to the host galaxy's fitted center to be the jet. We used \texttt{PyBDSF} to calculate the PA and center of the source. 

We aimed to always report the PA of the innermost jet. In some cases, \texttt{PyBDSF} returned the PA of the larger-scale jet, so we instead measured the PA manually, centered on the host galaxy's fitted center. We note, however, that our PA measurements cannot account for cases where the innermost jet bends on scales below the resolution of the radio survey used. We indicate our measured jet PAs with orange lines in Fig. \ref{fig:Obs_gallery_1}.

\section{Results} \label{sec:results}

In Fig. \ref{fig:Obs_gallery_1}, we show an optical and radio image of each galaxy in our sample, with the measured PAs and other relevant information indicated. We sorted the morphology of each galaxy's dust feature by visual inspection into one of four categories: disk, ring, lane, or unclear. We reported the angular and physical size of each image. To convert from angular to physical scales, we obtained the distance to each galaxy from NED. In cases where NED did not have a redshift-independent distance available, we used the galaxy's redshift, and calculated the angular diameter distance assuming cosmological parameters from \cite{Planck2018}.

We present our results in tabular form in Table \ref{tab:properties}. In addition to the PAs measured in this work, we give PAs from 6 previous papers that studied dust in HST-imaged radio galaxies. All of them included $PA_{\rm dust}$ and $PA_{\rm jet}$, but \cite{Capetti+2000} and \cite{deKoff+2000} did not include $PA_{\rm gal}$. \cite{Schmitt+2002} took many of their PAs from previous works; in cases where their PAs were taken from one of the other five papers referenced here, we excluded their values in order to avoid repetition. Note that all PAs are given as values between $0^\circ$ and $180^\circ$, measured from north through east.

\begin{figure}
    \centering
    \includegraphics[width=0.85\columnwidth]{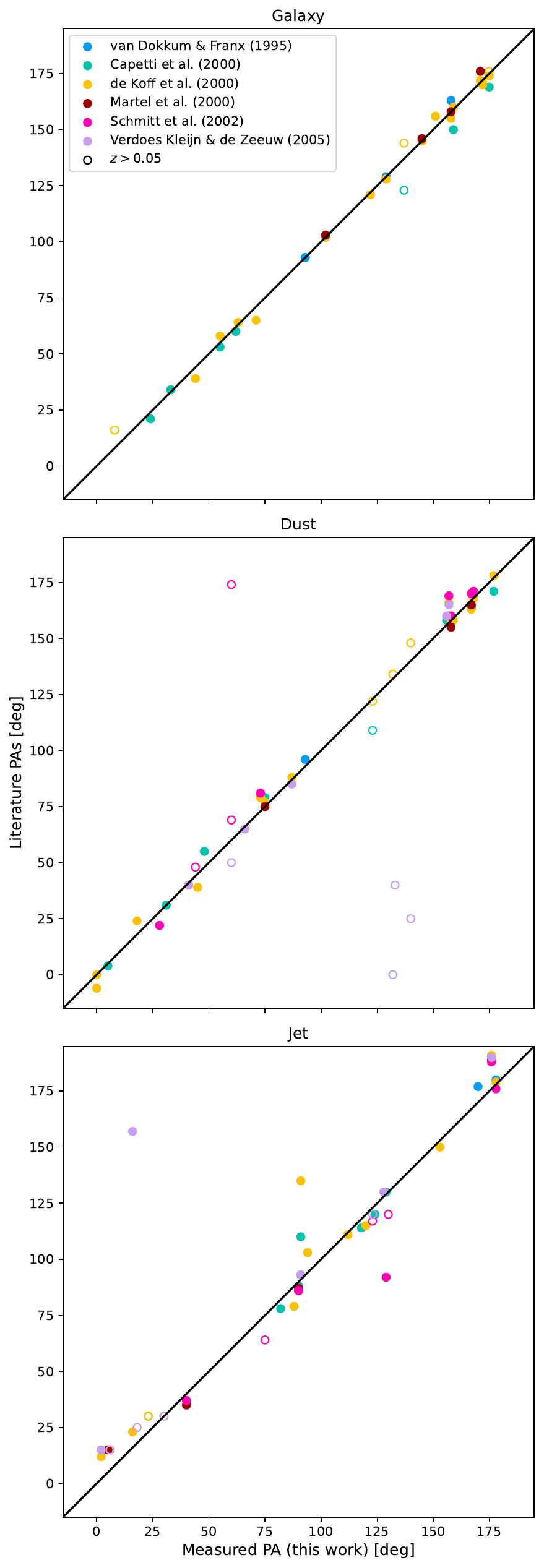}
    \caption{Galaxy, dust, and jet PAs from previous papers compared to the PAs measured in this work. The black line marks equality between the literature and measured PAs. The open markers indicate galaxies that fall on the high end of our covered redshift range. For points with literature PA $<15^\circ$ and measured PA $>165^\circ$, or vice versa, we shifted the literature PA by $\pm 180^\circ$ to simplify visual comparison with the measured PA.}
    \label{fig:PA_comparison}
\end{figure}

In Fig. \ref{fig:PA_comparison}, we show our measured PAs compared to the literature PAs, and find that they are mostly consistent. In cases where a literature PA is significantly different from our measurement, there is usually another literature PA for the same galaxy that is in closer agreement with our value. As such, we use our measurements alone for the analysis in Sec. \ref{sec:analysis}.

In Table \ref{tab:properties} we report the redshift of each galaxy from NED. The dust disks all fall on the low end of the redshift range covered by our sample, likely because they are more compact and therefore more difficult to observe at large distances. Fig. \ref{fig:PA_comparison} shows that the higher-redshift dust features often have more variance between our measured PAs and the literature values, but we find that the measured PAs of these features are still generally consistent with their lower-redshift counterparts.

\begin{deluxetable*}{c c c c c c c c c}
\tablehead{
Galaxy & Redshift & Dust & $PA_{\rm dust}$ & $PA_{\rm gal}$ & $PA_{\rm jet}$ & $PA_{\rm dust}$ & $PA_{\rm gal}$ & $PA_{\rm jet}$ \\
name & & morphology & (this work) & (this work) & (this work) & (literature) & (literature) & (literature)}
\startdata
3C 76.1 & $0.032489$ & Disk & $45^\circ$ & $129^\circ$ & $112^\circ$ & $39^\circ$ [6] & $128^\circ$ [6] & $111^\circ$ [6] \\
3C 236 & $0.1005$ & Ring & $60^\circ$ & $52^\circ$ & $123^\circ$ & $50^\circ$ [2], $174^\circ$ [3] & --- & $120^\circ$ [2], $117^\circ$ [3] \\
3C 293 & $0.045194$ & Lane & $28^\circ$ & $83^\circ$ & $129^\circ$ & $22^\circ$ [3] & --- & $92^\circ$ [3] \\
3C 321 & $0.096456$ & Lane & $60^\circ$ & $112^\circ$ & $130^\circ$ & $69^\circ$ [3] & --- & $120^\circ$ [3] \\
3C 403 & 0.060262 & Ring & $44^\circ$ & $34^\circ$ & $75^\circ$ & $48^\circ$ [3] & --- & $64^\circ$ [3] \\
B2 0908+37 & $0.104$ & Unclear & $133^\circ$ & $87^\circ$ & $6^\circ$ & $40^\circ$ [2] & --- & $15^\circ$ [2] \\
B2 0915+32 & $0.06099$ & Unclear & $123^\circ$ & $137^\circ$ & $23^\circ$ & $109^\circ$ [5], $122^\circ$ [6] & $123^\circ$ [5], $144^\circ$ [6] & $30^\circ$ [5], $30^\circ$ [6] \\
B2 1346+26 & $0.063309$ & Lane & $132^\circ$ &$8^\circ$ & $18^\circ$ & $0^\circ$ [2], $134^\circ$ [6] & $16^\circ$ [6] & $25^\circ$ [2], $25^\circ$ [6] \\
IC 4296 & $0.012465$ & Disk & $75^\circ$ & $62^\circ$ & $129^\circ$ & $79^\circ$ [5] & $60^\circ$ [5] & $130^\circ$ [5] \\
MCG-02-36-02 & $0.036689$ & Unclear & $48^\circ$ & $55^\circ$ & $124^\circ$ & $55^\circ$ [5] & $53^\circ$ [5] & $120^\circ$ [5] \\
NGC 193 & $0.014723$ & Lane & $0^\circ$ & $55^\circ$ & $94^\circ$ & $0^\circ$ [6] & $58^\circ$ [6] & $103^\circ$ [6] \\
NGC 315 & $0.016485$ & Disk & $41^\circ$ & $44^\circ$ & $128^\circ$ & $40^\circ$ [2], $40^\circ$ [6] & $39^\circ$ [6] & $130^\circ$ [2], $130^\circ$ [6] \\
NGC 547 & $0.018239$ & Unclear & $75^\circ$ & $102^\circ$ & $5^\circ$ & $75^\circ$ [4], $77^\circ$ [6] & $103^\circ$ [4], $102^\circ$ [6] & $15^\circ$ [4], $15^\circ$ [6] \\
NGC 612 & $0.030478$ & Ring & $175^\circ$ & $179^\circ$ & $100^\circ$ & --- & --- & --- \\
NGC 708 & $0.015886$ & Lane & $169^\circ$ & $89^\circ$ & $85^\circ$ & --- & --- & --- \\
NGC 1265 & $0.025665$ & Disk & $168^\circ$ & $172^\circ$ & $90^\circ$ & $171^\circ$ [3], $168^\circ$ [6] & $170^\circ$ [6] & $86^\circ$ [3], $86^\circ$ [6] \\
NGC 2329 & $0.01933$ & Disk & $0^\circ$ & $172^\circ$ & $153^\circ$ & $174^\circ$ [6] & $171^\circ$ [6] & $150^\circ$ [6] \\
NGC 3557 & $0.01027$ & Disk & $31^\circ$ & $33^\circ$ & $82^\circ$ & $31^\circ$ [5] & $34^\circ$ [5] & $78^\circ$ [5] \\
NGC 3801 & $0.011064$ & Lane & $18^\circ$ & $122^\circ$ & $120^\circ$ & $24^\circ$ [6] & $121^\circ$ [6] & $115^\circ$ [6] \\
NGC 4061 & $0.02445$ & Ring & $175^\circ$ & $174^\circ$ & $90^\circ$ & --- & --- & --- \\
NGC 4261 & $0.007261$ & Disk & $167^\circ$ & $158^\circ$ & $90^\circ$ & $164^\circ$ [1], $170^\circ$ [3], & $163^\circ$ [1],  & $88^\circ$ [1], $86^\circ$ [3], \\
& & & & & & $165^\circ$ [4], $163^\circ$ [6] & $158^\circ$ [4], $155^\circ$ [6] & $87^\circ$ [4], $87^\circ$ [6] \\
NGC 4335 & $0.015371$ & Ring & $159^\circ$ &  $151^\circ$ & $88^\circ$ & $158^\circ$ [6] & $156^\circ$ [6] & $79^\circ$ [6] \\
NGC 4374 & $0.003392$ & Lane & $73^\circ$ & $129^\circ$ & $178^\circ$ & $80^\circ$ [1], & $129^\circ$ [1], $128^\circ$ [6] & $0^\circ$ [1], \\
& & & & & & $81^\circ$ [3], $79^\circ$ [6] & & $176^\circ$ [3], $179^\circ$ [6] \\
NGC 4869 & $0.02282$ & Unclear & $177^\circ$ & $175^\circ$ & $91^\circ$ & $171^\circ$ [5], $178^\circ$ [6] & $169^\circ$ [5], $174^\circ$ [6] & $110^\circ$ [5], $135^\circ$ [6] \\
NGC 5141 & $0.017337$ & Lane & $87^\circ$ & $71^\circ$ & $2^\circ$ & $85^\circ$ [2], $88^\circ$ [6] & $65^\circ$ [6] & $15^\circ$ [2], $12^\circ$ [6] \\
NGC 5322 & $0.005937$ & Disk & $93^\circ$ & $93^\circ$ & $170^\circ$ & $96^\circ$ [1] & $93^\circ$ [1] & $177^\circ$ [1] \\
NGC 5532 & $0.02471$ & Disk & $158^\circ$ & $145^\circ$ & $40^\circ$ & $160^\circ$ [3], & $146^\circ$ [4], $145^\circ$ [6] & $37^\circ$ [3], \\
& & & & & & $155^\circ$ [4], $157^\circ$ [6] & & $35^\circ$ [4], $37^\circ$ [6] \\
NGC 6251 & $0.02471$ & Disk & $5^\circ$ & $24^\circ$ & $118^\circ$ & $4^\circ$ [5] & $21^\circ$ [5] & $114^\circ$ [5] \\
NGC 7052 & $0.015584$ & Ring & $66^\circ$ & $63^\circ$ & $16^\circ$ & $65^\circ$ [2], $65^\circ$ [6] & $64^\circ$ [6] & $157^\circ$ [2], $23^\circ$ [6] \\
UGC 367 & $0.03173$ & Disk & $156^\circ$ & $159^\circ$ & $91^\circ$ & $160^\circ$ [2], & $150^\circ$ [5], $160^\circ$ [6] & $93^\circ$ [2], \\
& & & & & & $158^\circ$ [5], $160^\circ$ [6] & & $93^\circ$ [5], $93^\circ$ [6] \\
UGC 9861 & $0.065525$ & Unclear & $140^\circ$ & $175^\circ$ & $30^\circ$ & $25^\circ$ [2], $148^\circ$ [6] & $176^\circ$ [6] & $30^\circ$ [2], $30^\circ$ [6] \\
UGC 12064 & $0.016694$ & Ring & $157^\circ$ & $171^\circ$ & $176^\circ$ & $165^\circ$ [2], $169^\circ$ [3], & $176^\circ$ [4], $172^\circ$ [6] & $10^\circ$ [2], $8^\circ$ [3], \\
& & & & & & $160^\circ$ [4], $166^\circ$ [6] & & $9^\circ$ [4], $11^\circ$ [6] \\
\enddata
\caption{Properties of the galaxies in our sample. References for the PAs from the literature are as follows: [1] \cite{vanDokkum_Franx_1995}; [2] \cite{Capetti+2000}; [3] \cite{deKoff+2000}; [4] \cite{Martel+2000}; [5] \cite{Schmitt+2002}; [6] VK05.}
\label{tab:properties}
\end{deluxetable*}

\begin{figure*}
    \centering
    \includegraphics[width=2\columnwidth]{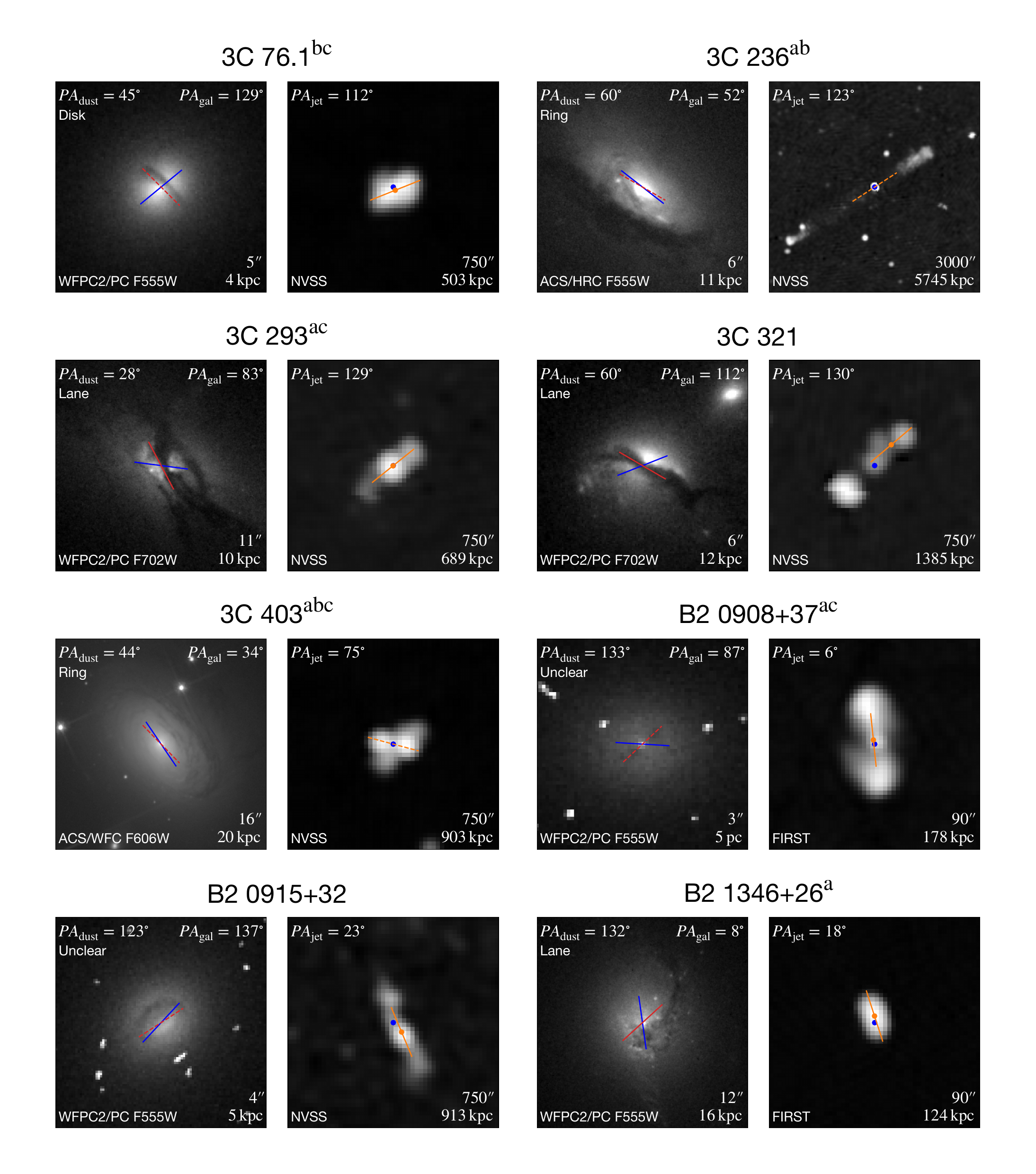}
    \caption{Image gallery of the galaxy sample, Part 1/4. We applied an asinh scaling to the images for visualization. For each galaxy, we show the optical HST image on the left and the radio image on the right. \textbf{Optical image:} We report the measured dust and galaxy PAs, the dust morphology, the HST instrument and filter used, and the angular and physical side length of the box. We mark the dust and galaxy PAs in the image with red and blue lines, respectively, centered on the galaxy position. A solid red line indicates an automatic dust fit, while a dashed red line indicates a manual fit. \textbf{Radio image:} We report the measured jet PA, the survey used, and the angular and physical side length of the box. We mark the center of the galaxy with a blue point and the center of the jet with an orange point. A solid orange line indicates an automatically measured jet PA, while a dashed orange line indicates a manual measurement. \textbf{Superscripts:} These indicate the following adjustments: (a/b) In the optical/radio image, we adjusted the upper limit of the color map, the linear width parameter of the asinh scaling, or both; (c) We cut the optical image to avoid a bright spot or defect; (d) We used the second optical image in the observational table rather than the first.}
    \label{fig:Obs_gallery_1}
\end{figure*}

\begin{figure*}
    \ContinuedFloat
    \centering
    \includegraphics[width=2\columnwidth]{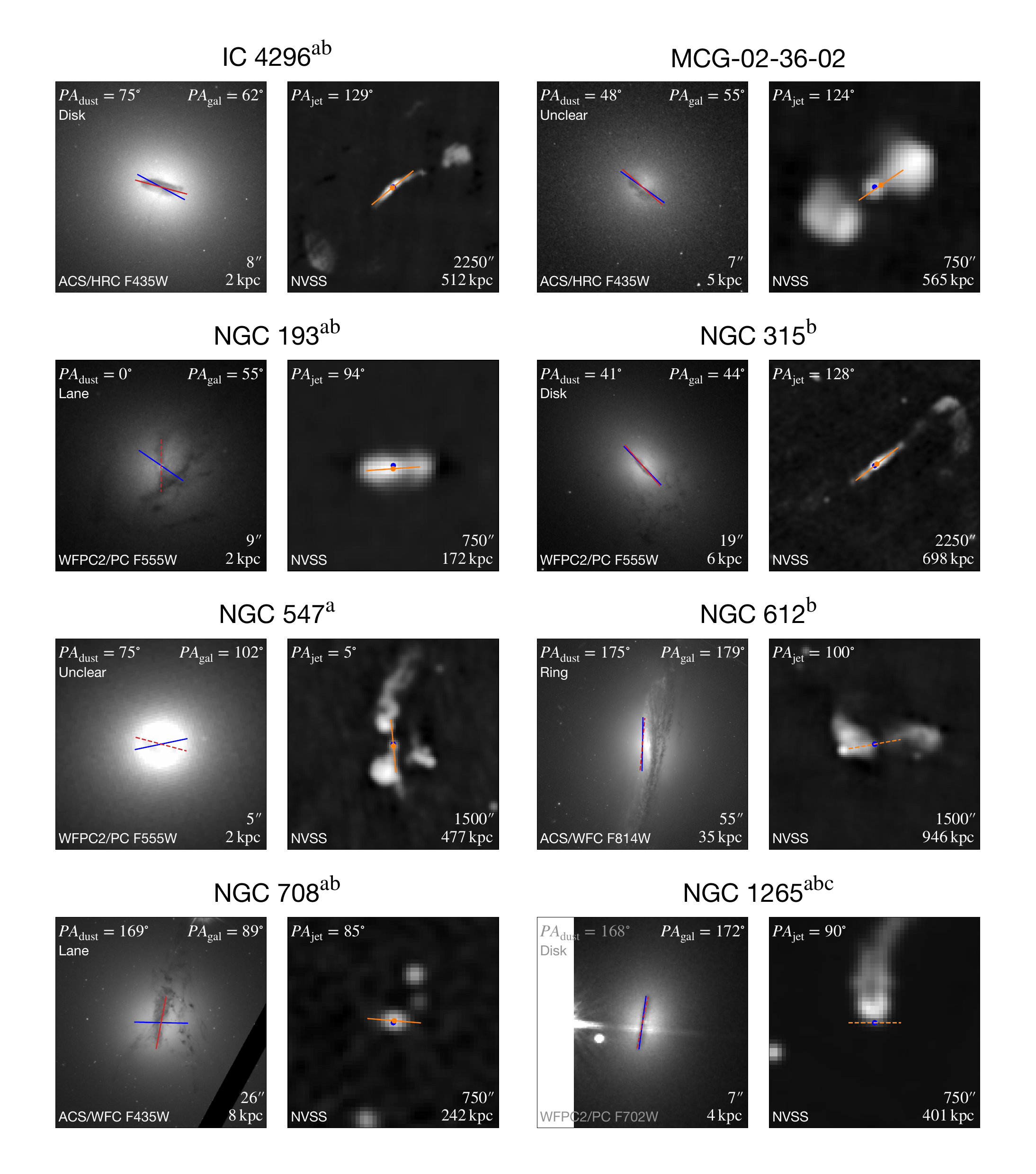}
    \caption{Image gallery of the galaxy sample, Part 2/4.}
    \label{fig:Obs_gallery_2}
\end{figure*}

\begin{figure*}
    \ContinuedFloat
    \centering
    \includegraphics[width=2\columnwidth]{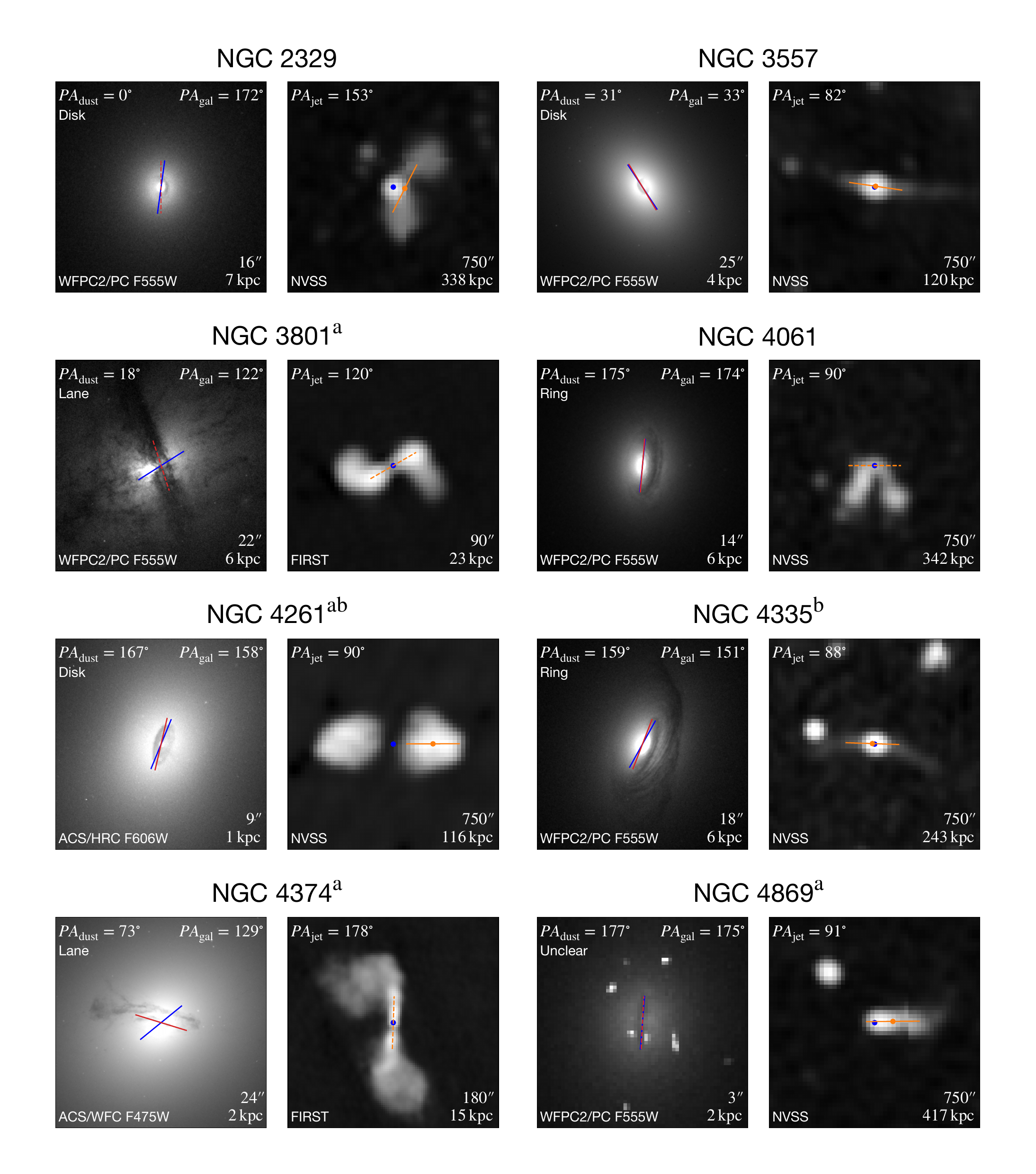}
    \caption{Image gallery of the galaxy sample, Part 3/4.}
    \label{fig:Obs_gallery_3}
\end{figure*}

\begin{figure*}
    \ContinuedFloat
    \centering
    \includegraphics[width=2\columnwidth]{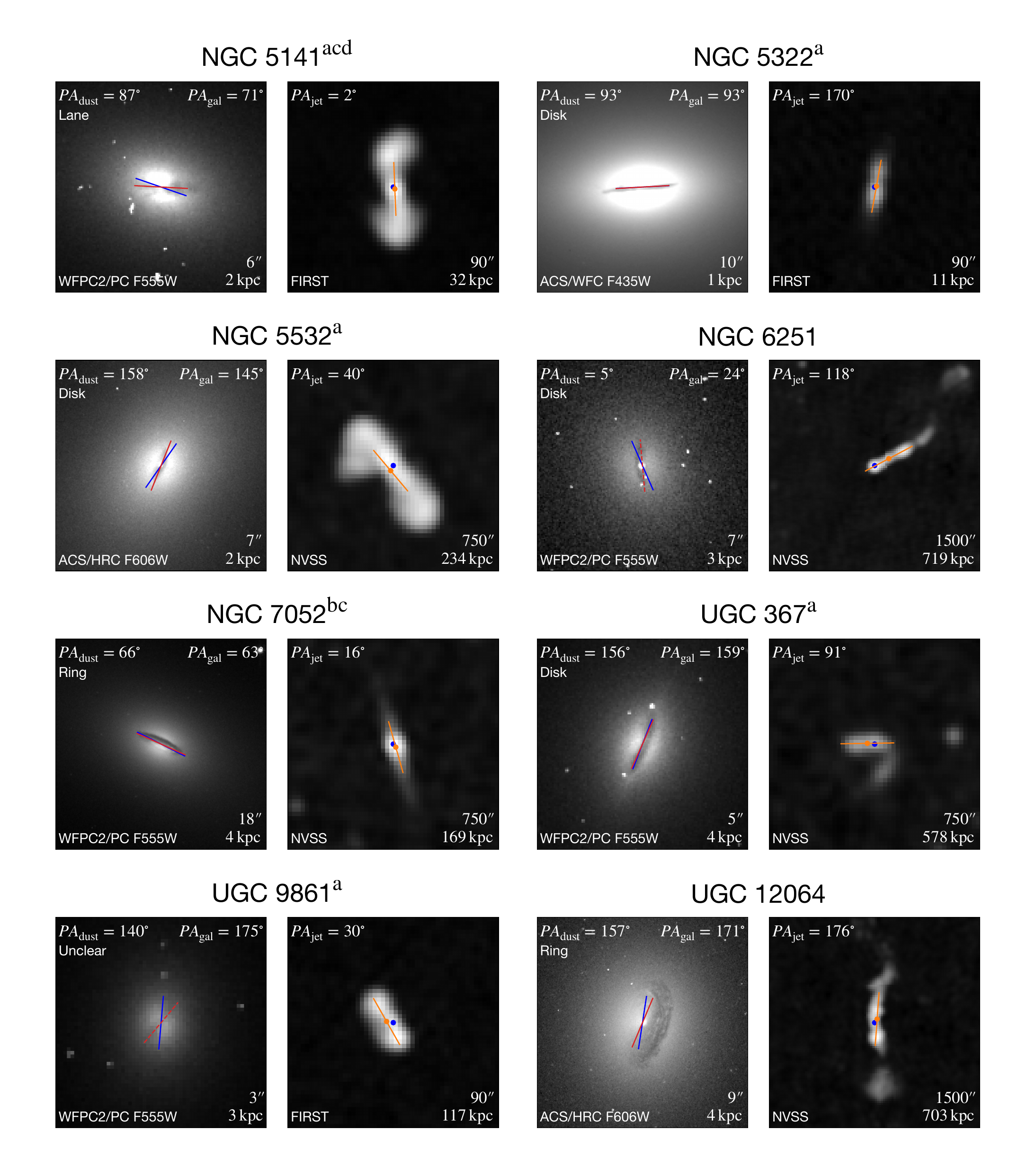}
    \caption{Image gallery of the galaxy sample, Part 4/4.}
    \label{fig:Obs_gallery_4}
\end{figure*}

\newpage

\section{Analysis} \label{sec:analysis}

From the PAs for each galaxy, we calculate $\padiffdustjet$, the projected angle between the jet and the major axis of the dust feature, which we require to be between $0^\circ$ and $90^\circ$:
\begin{equation}
    \padiffdustjet = 
    \begin{cases}
        | PA_{\rm dust} - PA_{\rm jet} |, \text{ if this is} \leq 90^\circ \\
        180^\circ - | PA_{\rm dust} - PA_{\rm jet} |, \text{ else}
    \end{cases}
\end{equation}
We also calculate $\padiffgaljet$, the projected angle between the jet and the galaxy major axis, and $\padiffgaldust$, the projected angle between the major axes of the dust and the galaxy.

We note that we are dealing with projected two-dimensional angles. For discussion of the intrinsic, three-dimensional distribution of galaxy-dust-jet alignment, see, for example, \cite{vanDokkum_Franx_1995}, \cite{deKoff+2000}, \cite{Schmitt+2002}, and \cite{VerdoesKleijn_deZeeuw_2005}. See also \cite{Ruffa+2020} for a three-dimensional analysis of the relative orientation of CO disks and radio jets in individual galaxies.

\begin{figure}
    \centering
    \includegraphics[width=\columnwidth]{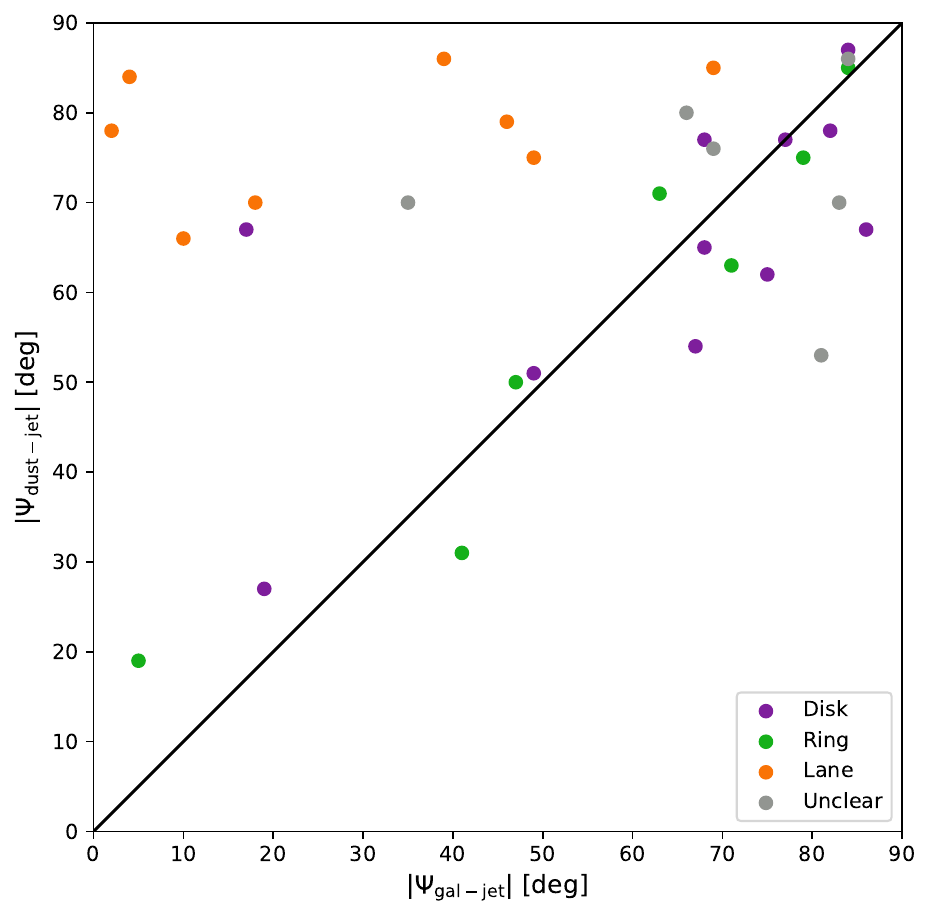}
    \caption{$\padiffdustjet$ vs. $\padiffgaljet$ for the galaxies in our sample. The black line marks $\padiffdustjet = \padiffgaljet$. The points are color coded by dust morphology.}
    \label{fig:PA_scatter_this-work}
\end{figure}

\begin{figure}
    \centering
    \includegraphics[width=0.9\columnwidth]{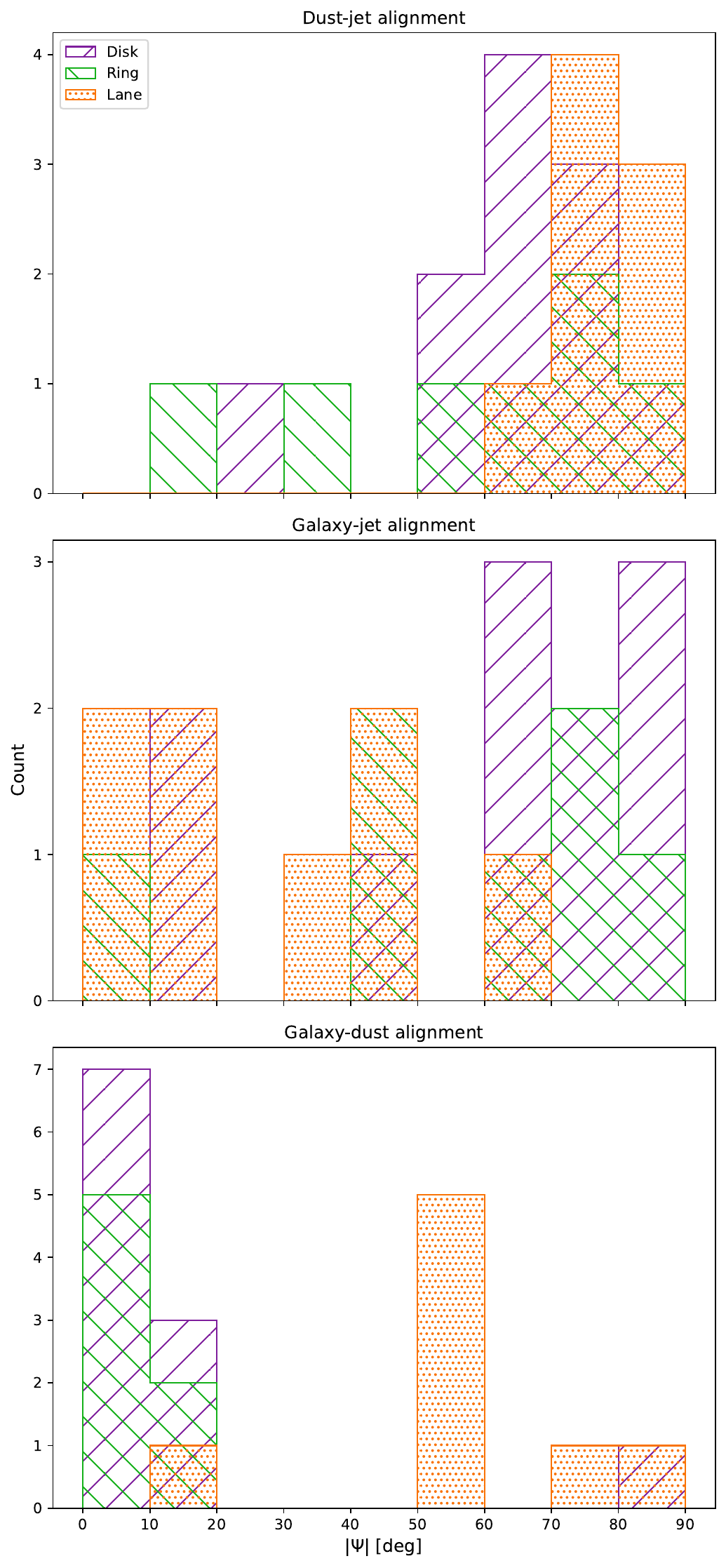}
    \caption{Distribution of $\padiffdustjet$, $\padiffgaljet$, and $\padiffgaldust$ for the galaxies in our sample, separated by dust morphology.}
    \label{fig:PA_hist}
\end{figure}

In Fig. \ref{fig:PA_scatter_this-work}, we show $\padiffdustjet$ vs. $\padiffgaljet$ for the galaxies in our sample. We break down these results in Fig. \ref{fig:PA_hist} with histograms of $\padiffdustjet$, $\padiffgaljet$, and $\padiffgaldust$, separated by dust morphology. We compared the distributions to a uniform distribution  using Kolmogorov–Smirnov (KS) tests.

We see that the dust lane galaxies have large $\padiffdustjet$ ($p = 3.2 \times 10^{-5}$ for a one-sided KS test), with a minimum of $66^\circ$. $\padiffgaldust$ varies for these galaxies, but tends towards larger values ($p = 2.5 \times 10^{-2}$ for a one-sided KS test). These findings are consistent with VK05. They found that jets are preferentially perpendicular to dust lanes, and that lanes do not have a preferred alignment with respect to their host galaxies. The exception is large ($>{\rm kpc}$) lanes, which are preferentially perpendicular to the galaxy major axis. This may explain why we see an overabundance of high $\padiffgaldust$ values, as we include many large dust lanes. The distribution of $\padiffgaljet$ is consistent with a uniform distribution ($p=0.28$ for a two-sided KS test).

Again in agreement with VK05, we find that dust disks and rings (equivalent to VK05's dust ellipses) are closely aligned with the major axes of their host galaxies. $\padiffgaldust < 20^\circ$ with one exception (3C 76.1), and the distributions are skewed towards low values ($p = 3.4 \times 10^{-6}$ for disks and $p = 2.2 \times 10^{-6}$ for rings, for one-sided KS tests). The jet alignment varies. For dust rings, the distributions of $\padiffdustjet$ and $\padiffgaljet$ are consistent with a uniform distribution ($p = 0.59$ and $p = 0.42$, respectively, for two-sided KS tests).

For dust disks, $\padiffdustjet$ and $\padiffgaljet$ also cover a range of values, but many of the galaxies are somewhat clustered in the upper right corner of the $\padiffdustjet$ vs. $\padiffgaljet$ scatterplots. The distributions of $\padiffdustjet$ and $\padiffgaljet$ are skewed towards high values ($p = 4.1 \times 10^{-3}$ and $4.6 \times 10^{-3}$, respectively, for one-sided KS tests). These high $\padiffgaljet$ values support the findings of \cite{Zheng+2024}. The lack of such a trend for dust \textit{lane} galaxies is also in line with their work: they suggested that there is no preferential galaxy-jet alignment for galaxies that have experienced recent mergers.

As mentioned in VK05, the difference between dust lanes, disks, and rings may explain the varying results of previous studies on dust-jet alignment: those that primarily included dust lanes would find a preferentially perpendicular alignment, while those that focused on well-defined disk/ring structures would not see this trend.

\section{Discussion} \label{sec:discussion}

VK05 proposed two possible explanations for the difference in alignment behavior between dust lanes and ellipses: (i) Dust lanes are formed when a jet produces a torque in the surrounding hot X-ray gas, forcing the dusty ISM into a plane perpendicular to the jet. (ii) The dust originates from outside the galaxy. The jet turns on and is initially aligned with the angular momentum of the infalling dusty material, but the dust lane eventually settles into an ellipse and loses its relation with the jet orientation.

As mentioned in Sec. \ref{sec:intro}, \cite{Kaviraj+2012} and \cite{Shabala+2012} found several indicators that that dust lanes in early-type galaxies originate from gas-rich minor mergers. The lanes tend to be morphologically disturbed and several times more massive than the maximum amount expected from stellar mass loss. Even after accounting for younger starburst ages, early-type galaxies with dust lanes exhibit higher than average star formation and black hole accretion rates, consistent with them having experienced recent gas-rich mergers. This suggests that the first explanation offered by VK05 is not the full picture. The dust disks and rings may be the settled successors of dust lanes, as proposed in the second explanation from VK05. Alternatively, they may originate internally via mechanisms such as stellar mass loss and chaotic cold accretion (e.g. \citealt{Gaspari+2013, Olivares+2022}).

The mechanism by which the jet becomes aligned with infalling material has been studied theoretically. For example, \cite{Natarajan_Pringle_1998} and \cite{Natarajan_Armitage_1999} show that the torque that aligns the inner accretion disk with the SMBH, known as the Bardeen-Petterson effect, also aligns the spin of the SMBH with the angular momentum of the outer accretion disk. This occurs on a timescale much shorter than the age of the jet. They therefore suggest that the jet direction is determined by the angular momentum of the accreted material or by the gravitational potential of the host galaxy. However, if an SMBH has a low accretion rate and thick accretion disk, which is likely the case for most of our radio galaxies, then the Bardeen-Petterson effect does not apply. The alignment is slower, and primarily affects only the inner disk \citep{Natarajan_Armitage_1999, McKinney+2013}. 

Even in cases where the SMBH spin is not aligned and the jet is launched into the dust feature, it would likely destroy the dust (VK05) or be deflected. This can explain the near-absence of galaxies with $\padiffdustjet < 20^\circ$.

\begin{figure*}
    \centering
    \includegraphics[width=2\columnwidth]{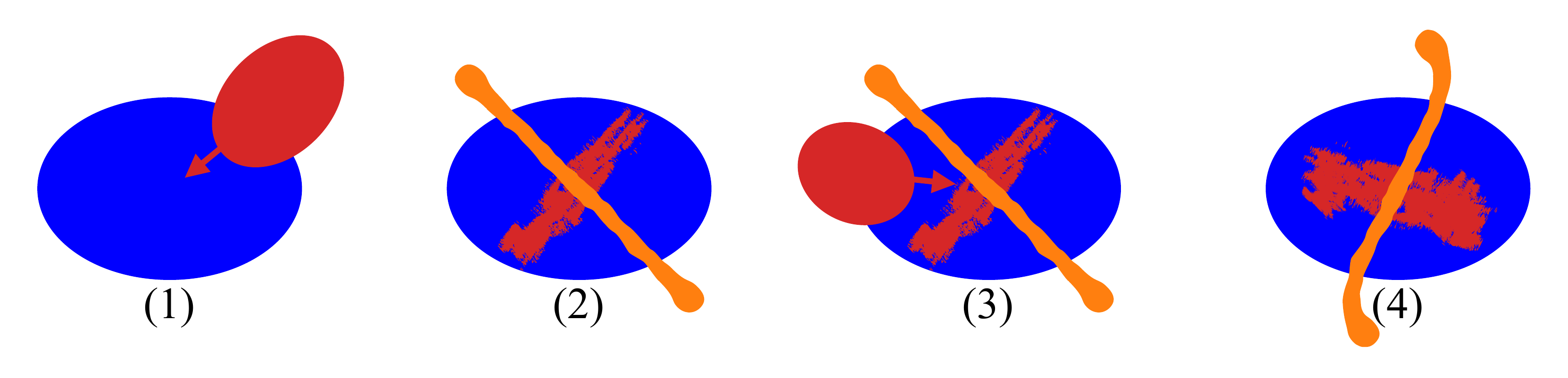}
    \caption{We illustrate a possible process by which galaxy mergers impact the jet angle: (1) An early-type galaxy experiences a merger with a smaller, gas-rich galaxy. (2) The early-type galaxy now contains gas and dust from the merger and features a prominent dust lane. The infalling gas and dust feed the central SMBH and the AGN turns on, creating a jet perpendicular to the dust lane. (3) The galaxy experiences another minor merger. (4) The second merger produces a dust lane in a different orientation, and the jet angle changes.}
    \label{fig:cartoon}
\end{figure*}

Putting this all together, we propose the following picture, illustrated in Fig. \ref{fig:cartoon}: An early-type galaxy experiences a gas-rich minor merger, bringing in gas and dust at a random angle and forming a dust lane. This triggers an AGN radio jet (if the jet is not already on) that quickly aligns with the angular momentum of the infalling material. Subsequent mergers create dust lanes in different directions, causing the jet angle to change.

This picture has important implications for halo heating and galaxy quenching. If correct, it means that the radio jet orientation changes each time a minor merger brings in new gas and dust. While these events can trigger starbursts in the short term, in the long term they allow for more isotropic heating of the ISM than if the jet maintained a constant orientation through cosmic time, and may help explain how massive galaxies maintain quiescence.

Many of the galaxies in our sample have bent jets, as seen in Fig. \ref{fig:Obs_gallery_1}. Some of these may represent cases in which the jet was already on but experienced a change in angle due to a galaxy merger, as in the final panel of Fig. \ref{fig:cartoon}.

While not illustrated in Fig. \ref{fig:cartoon}, dust disks and rings likely form from settling dust lanes, stellar mass loss, or chaotic cold accretion, as mentioned above. If they have both internal and external formation mechanisms, this might explain the observed dichotomy in dust disk alignment, where some are clustered at high $\padiffdustjet$ and $\padiffgaljet$ while others cover the full range of alignment angles.

\section{Conclusions} \label{sec:conclusions}

In this work, we followed up on previous studies of galaxy-dust-jet alignment. We took archival HST optical imaging and VLA radio imaging of early-type radio galaxies containing nuclear dust, and designed a largely automated process for measuring PAs. We found that dust lanes, which likely originate from gas-rich minor mergers, do not have a preferred alignment relative to their host galaxies. However, most are approximately perpendicular to the radio jets. In contrast, dust disks and rings are typically closely aligned with the major axes of their host galaxies, but their alignment with the jet varies. Our results suggest that infalling gas and dust from mergers help determine the angle of the resulting AGN radio jet. This means that the jet orientation would change as a result of mergers, heating the halo over a range of angles. As such, the connection between dust features and radio jets may have important implications for the role of AGNs in maintaining quiescence in massive early-type galaxies.

\begin{acknowledgements}
    We thank the anonymous referee for their excellent report that improved the paper.
    We thank Julie Hlavacek-Larrondo, Lena Murchikova, and Priyamvada Natarajan for helpful discussions.
    E.J.W. acknowledges support by the National Science Foundation Graduate Research Fellowship Program under Grant No. DGE-2139841. Any opinions, findings, and conclusions or recommendations expressed in this material are those of the authors and do not necessarily reflect the views of the National Science Foundation.
    This research is based on observations made with the NASA/ESA Hubble Space Telescope obtained from the Space Telescope Science Institute, which is operated by the Association of Universities for Research in Astronomy, Inc., under NASA contract NAS 5–26555.
    This research made use of NASA's \textit{SkyView} facility (\href{https://skyview.gsfc.nasa.gov}{skyview.gsfc.nasa.gov}), located at NASA Goddard Space Flight Center; the NASA/IPAC Extragalactic Database, which is funded by the National Aeronautics and Space Administration and operated by the California Institute of Technology; and Photutils, an Astropy package for detection and photometry of astronomical sources \citep{Bradley+2025}.
\end{acknowledgements}

\facilities{HST (ACS/HRC, ACS/WFC, WFPC2/PC), MAST, VLA}

\software{Astropy \citep{astropy:2013, astropy:2018, astropy:2022}, Astroquery \citep{Ginsburg_2019}, Matplotlib \citep{Hunter_2007}, NumPy \citep{Harris+2020}, Photutils \citep{Bradley+2025}, PyBDSF \citep{Mohan_Rafferty_2015}, SciPy \citep{Virtanen+2020}}

\bibliography{main}{}
\bibliographystyle{aasjournalv7}

\end{document}